# Yellow Light Energy Transfer Emitting Diodes Based on mixed Quasi-2D Perovskites

Dionysios Papadatos, Anastasia Vassilakopoulou and Ioannis Koutselas*

Materials Science Department, School of Natural Sciences, University of Patras, Patras, 26504, Greece.

*ikouts@upatras.gr

**Abstract**

The new class of hybrid organic-inorganic semiconductor (HOIS) materials, based on halide perovskites, is constantly being pursued for applications such as Light Emitting Diodes (LEDs) and solar cells, due to their momentous optoelectronic properties. In this work, we present a single layer LED that operates due to energy transfer effects as well as a simple, instant and low cost method for its fabrication. A LED device based on a mixture of zero dimensional (0D) $(CH_3NH_3)_4PbI_6$, two dimensional (2D) $(F-C_6H_4CH_2CH_2NH_2)_2PbI_4$ and three dimensional (3D) $(CH_3NH_3)PbI_3$ HOIS, is presented for the first time. The final composite material manifests simple, yet unique energy transfer optical effects, while its electroluminescence exhibits excitonic recombination bright yellow light, peaked at 592 nm. LED device fabricated under ambient air, readily functions at room temperature and low voltages. As for the active layer, it exhibited substantial film continuity in any form of deposition. Finally, with appropriate mixtures, it is possible to create films containing phase changes that exhibit dual color emission, here presented as yellow-green.

**Introduction**

The monadic optoelectronic properties of methyl ammonium lead halides ($CH_3NH_3PbX_3$, X= I, Br, Cl),[1-7] as well as their related low cost and simple synthesis conditions, have rendered them as a point of interest for many years. As a direct consequence, Light Emitting Diodes (LEDs) and solar cells have been fabricated from such semiconductor materials.[8-12] An interesting class of semiconductors which have revealed really interesting quantum phenomena in the past, are the Low Dimensional (LD) Hybrid Organic-Inorganic Semiconductors (HOIS).[13,14] Part of the impressive advantages of the LD HOIS is the high binding energy ($E_b$) and increased oscillator strength ($f_{exc}$) of their excitons, which result to useful tunable excitonic Optical Absorption (OA). In particular, the OA peak position can be tuned via altering halogen's stoichiometry or by simple synthetic path modification.[2,3] A HOIS is usually composed of a metal halide unit network, constituting the semiconducting inorganic part, while an electrooptically inactive amine takes the role of the organic part. HOIS materials synthesized by a broad range of inorganic and organic parameters, have led to various properties, such as tunable absorption in the whole ultraviolet-visible spectra,[14] energy transfer phenomena[15,16] or thin film transistor gate materials comparable to amorphous Si.[17,18] Finally, solar cell devices have been formed by three dimensional (3D) and two dimensional (2D) lead halide HOIS.[19,20] By altering the appropriate stoichiometry and synthetic route of both HOIS parts, it is possible to form semiconducting systems with dimensionality between 3D, one dimensional (1D) or even zero dimensional (0D) quantum dot like semiconductors.[2,21-27] Furthermore, HOIS where the active part has a dimensionality between 2D and 3D are possible to be synthesized, referred to as quasi-2D semiconductors. The presented self-assembled quasi-2D HOIS have a general chemical formula $(Z)_{n-1}(M-H)_2Pb_nX_{3n+1}$ where

n=1,2,3…, M is an amine[28-31] while M-H stands for the protonated amine, Z refers to CH$_3$NH$_3$ (Methylamine-H, abbr. MA-H) and X refers to I, Br, Cl. In this work, M will only be 4-Fluorophenethylamine (4FpA) and X=I. **Figure 1** shows structural patterns of these layered materials. When n=∞ (**Figure 1**, right) HOIS is equivalent to the 3D material (perovskite), while the material is strictly 2D when n=1 (**Figure 1**, left). In the latter case, organic cations are separating the active layers, each one being an infinite 2D sheet of edge sharing PbI$_6$ octahedra with thickness of ca. 6Å while layer's thickness is twice the length of the Pb-halogen bond. Inorganic layer's thickness increases linear with dimensionality's increase from 2D to 3D, approximately of ca. nx6Å, however, layer's stability is supported by MA-H cations.

The energy band gap ($E_g$) presents a reciprocal behavior of the quasi-2D HOIS dimensionality, thus, $E_g$ decreases as dimensionality increases, until it reaches that of the 3D HOIS. In addition, the excitonic $E_b$ which is depended on the organic molecule and the actual crystal structure, tends to decrease too.[2] In most cases, crystals of a specific crystalline quasi-2D HOIS are difficult to be obtained, especially for n>3. OA and Photoluminescence (PL) spectra in such systems, appear to be composed as sum of the corresponding spectra of the individual quasi-2D/3D crystalline HOIS.[32]

In this work we report a simple, instant and low cost method for the fabrication of a LED device constructed by a single active layer, composed of 0D, 3D and 2D HOIS blend. This blend containing specific (0D/3D):2D molar ratio, manifests simple, yet obvious energy transfer effects,[33] as these have been seen in semiconductor materials like CsPbBr$_3$,[34-36] while similar phenomena in LED have been observed in recent publications.[37,38] LED device, presented herein, can function at room temperature, ambient air and low voltages while the EL emission is visible by naked eye. The active layer showed strong film continuity and significant stability in conditions of high humidity. Finally, a set of simple experiments is presented in order to elucidate the exciton energy transfer model and the range of phenomena it can encompass, while for some experiments multi-color LED is reported.

**Experimental Details**

**Chemicals.** Gallium–indium eutectic (≥99.99% trace metals basis), Methylamine (Meth-40% in water), 4-Fluorophenethylamine (4FpA-99%), Indium tin oxide coated glass (ITO, square surface resistivity 15-25 Ω/sq), Acetonitrile (AcN-CHROMASOLV® Plus ≥99.9%), Hydriodic acid (HI-ACS reagent, ≥47.0%), Lead (II) iodide ($PbI_2$-99.999% trace metals basis), N,N'-Dimethylformamide (DMF-99.8%). All chemicals were obtained from Sigma-Aldrich and used without any further purification. All acid quantities described further in the text will refer to the previously mentioned respective solutions.

**Synthesis.** For the blend's synthesis containing 0D and 3D perovskites (MA-H)$_4$PbI$_6$.2H$_2$O and (MA-H)PbI$_3$, respectively, $PbI_2$ (1.6 mmol) was dissolved in 5 mL AcN with the addition of 0.170 mL of HI and stirred until an optically clear solution had been obtained. Methylamine (5 mmol) was dissolved as salt by stirring in 5 mL AcN with HI (0.301 mL). These two solutions were slowly mixed under room temperature, and as dried, a black precipitate was formed. Although, the 3D compound can be synthesized by using less methylamine, in this case, the surplus of methylamine leads to simultaneous formation of 3D and 0D crystalline units. The final blend will be denoted as zero and three dimensional (ZTD) semiconductor. A prediction of reaction's products was made using a non-linear fitting function in order to minimize the loss of precursors when these react to form 3D, 0D or other byproducts. This was performed assuming that the reactions have progressively consumed completely every precursor, while the 0D and 3D compounds can be each formed without any preferential choice by nature. We have estimated that the mole fraction of 0D to 3D in the final ZTD compound is 1.43:0.06. Within the same mathematical model low concentration of $PbI_2$ appears as unreacting which is, however, unobserved in the XRD patterns of ZTD. Therefore, the average molecular weight of the final ZTD material is estimated to be 1090 ± 20 g.

In the case of the 2D perovskite, $(4FpA-H)_2PbI_4$, $PbI_2$ (4 mmol) was dissolved in 5 mL AcN and after the addition of 0.425 mL of HI, it was stirred until an optically clear solution was obtained. 4-Fluorophenethylamine (8 mmol) was dissolved as salt by stirring in 5 mL AcN with HI (0.144 mL). These two solutions were slowly mixed under room temperature, yielding a bright orange precipitate. Slow cooling gave rise to red-orange crystals, which have been dried extremely well from solvent. The semiconductor shows strong PL even under room light.

In a typical preparation, the mixed semiconductors where prepared from a 1:1.4 molar ratio of the ZTD and 2D precursors. In particular, 30 mg ZTD and 49 mg $(4FpA-H)_2PbI_4$ were dissolved in 0.300 mL DMF, forming a yellow solution. The stirring temperature was set at 40 °C. This specific blend will be referred as **BY1**. This mixture is composed out of mainly three phases, as it will be discussed below, the n=1,2,3 quasi-2D HOIS. Again, without being able to analytically find the proportions of each phase, we used a nonlinear fitting method to find the best mathematical solution when the above mixture is completely consuming the precursors. The solution for the case described before appears to be for the mole ratio of $2D_{n=1}:2D_{n=2}:2D_{n=3}:2D_{n=4}:3D$ to be 2.11:0.29:0.13:0.12:0.08. Even if the mentioned analysis is strictly mathematical, it allows a quantitative pre-determination of the concentration of some of the quasi-2D phases. However, the model cannot discuss the arrangement of these phases, which are mainly nanosized entities. For the cases where twice the molar amount of either ZTD or 2D semiconductor was used, similar analysis results are partially portrayed in the experimental OA spectra and more specifically, in the position and intensity of the excitonic absorption peaks. Specifically, when each phase is abundant, a related strong excitonic peak is observed. However, when fast drying or equivalently high heating rates are used, the formation and concentration of each phase is a complex function of the solvent and the temperature that can only be experimentally determined.

**Device Fabrication.** LED prototype was created in a single layer fashion. ITO glass was coated with 0.200 mL of the BY1 solution, by two different processes. In the first process, the ITO glass was spin coated according to the following steps: i) 4000rpm for 20 sec, ii) 3000rpm for 20 sec, iii) deceleration in 20 sec. In the second process, the solution was left for drying onto glass's surface at the temperature of 80 ºC. Regardless the process, the film turned red and showed strong continuity.

ITO or FTO coated substrates were immersed in piranha solution for 15 min and rinsed with 18 Mohm water, however, such immersion is not crucial. In order to function the LED, a negative power supply was manually contacted with a Ga-In alloy tipped wire. Video showing LED's emission is provided in supplementary files.

**Characterization Techniques.** Powder X-ray diffraction (pXRD) data were obtained from polycrystalline samples on a Bruker D8 Advance diffractometer equipped with a LynxEye® detector and Ni filtered CuKa radiation. The scanning area covered the 2θ interval 2-80º, with a scanning angle step size of 0.015º and a time step of 0.161 sec.

OA spectra in the ultraviolet-visible spectral region were recorded on a Shimadzu 1650 spectrophotometer in the range of 200-800 nm, at a sampling step of 0.5 nm at 1.5 nm slits, using a combination of halogen and $D_2$ lamps as sources. Samples were measured as thin or thick films on quartz or ITO/FTO substrates, after having deducted the substrate's spectra as reference.

PL and Photoluminescence Excitation (PLE) spectra were obtained from solid pressed pellets or thick deposits on quartz plates, mounted on a Hitachi F-2500 FL spectrophotometer employing a xenon 150 W lamp and a R928 photomultiplier. The excitation and detection slits were set at 2.5-5.0 nm. For signals inducing saturation, lower accelerating voltage was chosen at 450 V, from 700 V. In some cases, PLE spectra have been corrected through the instrument supplied files. All OA, PL and PLE spectra were recorded at room temperature unless otherwise designated.

Electroluminescence spectra (EL) was recorded by modifying a θ-Metrisis optical profilometer equipped with an Ocean Optic polychromator CCD. LED used for EL characterization was composed of ITO glass, on which thick film of BY1 was deposited. Structural and spectroscopic information for the 2D HOIS semiconductor is obtained with instrumentation as before.[39] Electrical contact was manually made by rough copper wire covered with eutectic Ga-In alloy. Recorded spectra, video and images were easily visible by naked eye and recorded for forward bias only (ITO:+), of about 3-12 V, as reverse bias did not produce any light.

I-V characteristic curves were recorded using a Keithley 6517A measuring current at specified voltages with a sub-millimeter radius of Cu wire in the form of a spring slightly touching the surface of the active material. The small value of currents recorded in the I-V compared to those on by manual Ga-In contact, is probably due the area of the Ga/In liquid tip.

The morphology of the samples was observed by Scanning electron microscopy (SEM) (EVO-MA 10, Bruker) and Transmission electron microscopy (TEM) at 200 kV (JEOL, JEM 2100) where pictures were obtained with camera (Gatan).

**Results and Discussion**

A blend of 0D, 2D and 3D HOIS based on lead iodide, methylamine and 4-fluorophenethylamine, has been utilized for the first time. The reaction of (MA-H)$_4$PbI$_6$.2H$_2$O, (MA-H)PbI$_3$ and (4FpA-H)$_2$PbI$_4$ results to a blend of active HOIS materials, that each one exhibits low dimensional excitons with high binding energy, thus, leading the blend agglomerate to exhibit interesting quantum phenomena. The one of particular interest here is the energy transfer among excitons of neighboring such HOIS, giving rise to intense PL as well as intense EL at the agglomerate's moiety with the minimum band gap. Oddly enough, due to the previously mentioned effect and self-absorption of the composite films, it can be observed that films of the active materials show different PL originating from the front face (FF) and back face (BF) of the film, provided that illumination is only on one side. The respective PL on either side is naked-eye visible under UV excitation of 404 nm. As mentioned earlier, LED device using this blend was made with a simple and extremely low cost method that completes LED's fabrication in less than two minutes. LED device needed just one single active layer to operate, without the need of hole or electron injection layers.[26]

The XRD spectra presented in **Figure 2a**, shows PbI$_2$ pattern peaks as obtained from JCPDS-ICDD card 7−235, while the patterns of ZTD, 2D HOIS (4FpA-H)$_2$PbI$_4$ and their 0D/2D/3D blend BY1, can be seen in **Figures 2b**, **2c** and **2d**, respectively. It is first obvious that none of the spectra **2b**, **2c** or **2d** resembles uniquely that of PbI$_2$.

Specifically, the ZTD spectra (**Figure 2b**) consists mainly of a sum of the previously mentioned 0D and 3D XRD spectra. The spectra of the ZTD shows four strong peaks at 11.34º, 16.95º, 26.44º and 33.65º, probably due to the 0D compound, whose XRD spectra has been calculated from the respective crystal structure solution and is displayed in detail in the supplementary information (**Figure S1**).[27,29] As far as blend's 3D phase is concerned, some related minor peaks at 12.72º, 28.46º, 31.87º and 43.05º are exhibited. The first peak corresponds to the (001) $PbI_2$ plane while the second, third and fourth correspond to (220), (310) and (330) diffraction planes of (MA-H)$PbI_3$, respectively.[40] The above imply that the ZTD material is mainly composed of isolated octahedra, yet, as it will be shown later, its visual color implies the existence of a small fraction of 3D compound. In particular, the ZTD material presents a black color, however, when hand-pressed turns to yellow-like, while when re-dissolved in DMF and dried afterwards, red PL and EL can be observed. In time, the red EL glowing deposits may turn total black lacking red PL and EL, showing the intricate phenomena of this type of compounds which can acquire multiple structures, all equipped with strong excitonic states. **Figure 2c** is similar to other literature spectra,[41] while the spectra of the active material BY1 (**Figure 2d**) resembles that of other lately reported quasi-2D dimensional HOIS, which are analogous materials based on different amines.[29] Therefore, the XRD spectra in **Figure 2d** clearly shows the formation of a quasi-2D system, which does not enclose any of the ZTD or $PbI_2$ phases. Furthermore, the first peak corresponds to the formed bilayer (n=2) system, while the second peak at 5.78º corresponds to the monolayer (n=1) 2D HOIS. **Figures 2c** and **2d** show a partial resemblance since we have indeed used a 3D:2D molar ratio of 1:1.4 and it is, thus, expected to see the n=1 phase in the BY1 blend.

In order to investigate the optical properties of the perovskites used in this work, OA, PL, PLE and EL techniques were employed. **Figure 3** shows the OA spectra for ZTD (**3a**), BY1 (**3b**) and (4FpA-H)$_2$PbI$_4$ (**3c**). In **Figure 3a**, the excitonic OA peak of ZTD appears at 376 nm, which is related to the isolated, yet slightly interacting,[32] PbI$_6$ octahedra which themselves absorb at 365 nm, as it is also the case of DMF and AcN solutions of lead perovskite compounds.[32,42] The small red shift observed, for the 376 nm peak, is due to the blend of 0D with 3D perovskite, probably leading to significant interaction between the octahedra, thus, lowering their excitonic energy. **Figure 3b** shows the OA spectra of the BY1 blend. The first excitonic peak occurs at 518 nm indicating the formation of a single lead halide monolayer (n=1), sandwiched between organic layers. Similar excitonic peaks are observed for the (4FpA-H)$_2$PbI$_4$ compound (**Figure 3c**), where the excitonic binding energy, calculated from the onset of band gap absorption at 488 nm, is at least 116 meV. Also, a bilayer (n=2) lead halide HOIS has been formed in the BY1 film, which exhibits an excitonic peak at 568 nm. It is not easy to calculate excitonic binding energies for the n>1 species from the OA spectrum presented in **Figure 3b**, as it is composed of two different overlapping absorption spectra of 2D semiconductors. The previously mentioned optical absorption spectra, are also convoluted with 2D Sommerfeld factors, thus, their analysis would be only of mathematical importance at room temperature. In **Figure 3c**, a peak at 386 nm is also exhibited due to the PbI$_6$ octahedra energy levels perturbed by the adjoining octahedra, corresponding to the same peak as in the **Figure 3a** and **3b** spectra. No other excitonic peaks have been observed corresponding to n≥4, however, a small shoulder at ca. 610 nm can be discerned that is due to the formed trilayer (n=3) quasi-2D homologue material.[32] Peaks for n=1 and n=2 are quite distinct which proves that both HOIS have created a well crystalline blend. The concentration of each phase formed is an interesting topic, which could theoretically be calculated from the absorption coefficient, if the oscillator strength of each phase was taken into consideration. Although this is a subject for further research, it can be safely assumed that the number of n=2 nanoparticles in the BY1 blend is comparable to the number of nanoparticles of the n=1 phase. Other products will be detected in this spectral range characteristics, as the OA excitonic peaks of these two phases are very strong.

The PL spectra of ZTD ($\lambda_{exc}$=300 nm), BY1 ($\lambda_{exc}$=400 nm), (4FpA-H)$_2$PbI$_4$ ($\lambda_{exc}$=300 nm) as well of ZTD ($\lambda_{exc}$=400 nm) are shown in **Figure 4a**, **4b**, **4c** and **4d**, respectively. In general, the PL spectra exhibit excitonic recombination peaks that have a one to one correspondence to some OA excitonic peak, however, slightly red shifted. In detail, **Figure 4a** shows the PL spectra of the ZTD HOIS under 300 nm excitation, where the PL peaks appear at 340 nm and 500 nm of strong and weak intensity, respectively. The first peak is probably attributed to some high state of Pb-I interaction, while the second peak is related to radiative disexcitation of Pb$^{2+}$ ions.[43] It is, however, possible that some other phase or even PbI$_2$, yet undetectable in the XRD spectra, is also existing in ZTD exhibiting PL at ca. 500 nm, since normally the 500nm peak should be absent in a 0D HOIS.[32] The PL spectrum for the BY1 blend under 400 nm excitation is shown in **Figure 4b**, where peaks at 525 nm and 575 nm can be observed. For both peaks, a Stokes shift, w.r.t. to the OA excitonic peaks, of 7 nm is presented. The PL spectra of an ITO glass coated with (4FpA-H)$_2$PbI$_4$ under 300 nm excitation, is presented also in **Figure 4c**. The peak at 528 nm is associated with the free exciton recombination and as for the peak at 546 nm, is probably corresponding to either oxidized species or some defects inducing trapped excitons. The latter PL peak usually disappears once the material is recrystallized from DMF. **Figure 4d** shows the PL spectra of the ZTD HOIS under 400 nm excitation, where a weak PL peak at 461 nm is observed, which was less intense under 300 nm excitation.

PL peaks for each set of specific n quasi-2D nanoparticles detected in the OA spectra, are observed. It could be assumed that PL originating from a set of neighboring quantum entities, which is the geometrical arrangement of the BY1 sample, may have exhibited energy transfer phenomena, where energy would be transferred from the high energy band gap nanoparticles towards the lowest energy band gap nanoparticles in the blend.

In general, energy transfer in HOIS can be performed either by electric conduction of the carriers, i.e. electrons and holes, drifting towards the smallest band gap nanoparticles or by radiative coupling of high energy excitons formed by carrier attraction to existing nearby low energy excitonic states. Considering the BY1 OA peaks, this blend is indeed composed out of random arrangement of one 2D (n=1) and two quasi-2D (n=2,3) moieties. The lack of total energy transfer manifestation in its PL spectra, is probably indicating that the quantum entities for the n=2 or n=3 quasi-2D nanoparticles, due to their lower band gap and probable appropriate alignment of the bands,[37] cannot draw energy effectively from the n=1 nanoparticles' states. It is possible as well that the number of oscillators for n>1 is small, thus, energy transfer cannot completely take place. In the past, it has been observed that even if the OA of the lowest $E_g$ moiety, in a similar blend, is minimum, it is possible to have PL only at that lowest $E_g$ energy.[33] As a result, all energy observed was being transferred from the high energy states to the lowest, by virtue of Forster like energy transfer, where superradiance effects certainly must be taken into account as these link the excitonic wave functions along with the radiation field.[44,45] However, in this previous total energy transfer observation, the materials had been literally grinded together to form a set of well split and adjoined quantum entities.

On the contrary, here, the film has self-assembled to a set of three quantum entities, but mainly to the quasi-2D n=1,2,3 HOIS, which are not arranged in the best possible mixing geometry and be all at the same time in close proximity. This arrangement, contrary to the grinding method mentioned before, leads to the appearance of specific sharp OA peaks. Also, the small full width at half maximum (FWHM) of the respective OA excitonic peaks implies a well crystallized set of these various n nanoparticles. Moreover, both n=1 and n=2 nanoparticles may be rather large, ca. 20 nm, as evidenced by applying Scherrer's equation on the XRD spectra. Therefore, each large nanoparticle exhibits its own OA and PL spectra, while their agglomeration under specific geometrical arrangement conditions may not exhibit collective emission effects. Such effects are usually evidenced by bright PL that takes place when the exciton/radiation coupling does not loose coherence while travelling for distances comparable to the particles' diameter and that demands appropriate particle sizes.[35]

**Figure 5a** shows the PL spectra of a BY1 layer deposited on quartz, excited under UV radiation with BF configuration, where the radiation excites the same face where the originating PL is being measured from. In order to test for any phase separation that would have occurred along the film and perpendicular to the surface, the same layer was measured while radiation was exciting the other side of the sample, **Figure 5b**. It is obvious that the spectra presented in **Figure 5a** and **5b** are almost identical, thus, the crystallization on the ITO interface is the same as on the side in air, which shows that the material has uniformly dried throughout the sample volume as far as PL properties are concerned. On the other hand, there are differences between spectra of **Figure 5a** and **5b** to that of **5c**, which corresponds to a FF PL measurement. This spectrum portrays the absence of the green emission and only a red emission peak (606 nm) is observed, despite that the film is thinner from that corresponding to **5a** and **5b**. The latter peak is clearly red shifted by 114 meV from the lowest energy excitonic PL peak of BY1. This is probably attributed to self-absorption and re-emission, where emitting light wavelength is being led towards the moieties with defects states or very small band gaps, as it propagates in the film.

**Figure 6** shows the uncorrected PLE spectra for (a) BF and (b) FF configurations of a thick BY1 film, observed at 650 nm emission, while it is excited on the back face. It is logical to notice that the FF emission in the deep red is better observed when excited at 535 nm, with respect to the BF observation, since in this manner the whole depth of the film participates in the mentioned effects of self-absorption and re-emission, leading to the deep energy levels luminescence on the back of the film. The peak 535 nm should have normally been detected at the OA peak, ca. 518 nm, corresponding to that of the first exciton, however, it appears that again re-absorption phenomena may be responsible for this shift. The other peaks observed are related to the Xe lines used by PL spectrometer's lamp.

**Figure 7** shows the EL spectra of the BY1 blend, captured while LED device was emitting bright yellow naked-eye visible light, under 4-6 V turn on voltage. As a result, a distinct EL peak at 592 nm, is observed at all points on the film. This specific peak, is red shifted by 17 nm w.r.t. to the PL peak of the bilayer system, while no green EL arising from the monolayer 2D HOIS can be observed. The fact that no other wavelengths of emission are observed and that the peak is narrow, is a manifestation of the excitonic nature of the emission and that it assumed to be coupled to a specific quasi-2D crystalline HOIS. Image and video of LED's emission can be found as supplementary information. Finally, it should be stressed that LEDs based on the 4-FpA 2D (n=1) HOIS readily provide green light, which is not at all observed here. In the first level of analysis, the 3D HOIS has modified the network so that only the quasi-2D EL is observed. One would expect that since two phases exist, two emissions lines should also be detected. However, energy transfer phenomena are responsible for the lack of the 2D EL peak.

An I-V characteristic curve of a LED device based on BY1 blend is shown in **Figure 8**. It appears that for a voltage of about 5 V, a sharp decrease occurs in the current flowing, thus, for the emitted light intensity as well. This may be due to exciton diffusion and band bending phenomena,[46] but it is quite possible to be related to physicochemical changes occurring on the interface, either due to the Ga/In alloy, or some form of destruction on the contact point. In order to elucidate this point, more research is needed. The ideality factor of the diode by fitting the exponential part of the curve appears to be 76.

As further experimentations, an ITO substrate coated with BY1 at every point on its surface, was exposed for 48 hours under ambient air at 80 °C, in order to evaporate part of the methylammonium salt. In particular, after this period the material showed again visible-eye green PL from its BF, when excited with 404 nm radiation, while in thin areas PL appeared to be green in the FF configuration and red in regions of extremely thick film. The PL spectra of the heat treated sample is presented in **Figure 9**. The important point is that the 575 nm excitonic PL is hardly visible in the OA spectra, however, in all of the thin film regions the LED was turning on as yellow EL. Therefore, we immediately conclude that even if the excitonic peak for n=2 has vanished, the EL continues to appear from the almost diminished moieties of n=2, pointing to energy transfer. If there weren't any energy transfer effects, then one would assume that green EL should be observed, since the n=1 moiety appears to be dominant in the PL after heating. It is important to note again that the layers of (4FpA-H)$_2$PbI$_4$ do show green RT EL (555 nm) on their own, by simple touching of the Ga/In electrode on the active surface.[41] The red shift of 12 nm between the two spectra may be related to self-absorption and re-emission while the case of cavity effects has also been considered as a possible explanation.[46]

**Figure 10** presents BY1 sample's morphology, as captured using SEM (**Figure 10a-c**) and TEM (**Figure 10d-f**) techniques, without any gold coating in order to preserve as much as possible its surface topology. SEM images show that there is a distinct morphology of the sample that crystallized close to the ITO substrate as flakes, while at higher distances small rod-like crystals are observed. However, in all regions of flakes or rods, the molar ratio of I:Pb, as measured from EDX, is the same on both morphological features and close to 4, thus, the rods have no resemblance to PbI$_2$. The TEM images show that the composite film is composed of appearing white/grey areas, mixed with small size nanoparticles of the order of 30-50 nm, in accordance to the XRD analysis, thus, it is suspected that the quasi-2D are the black appearing nanoparticles, noted with red arrow in **Figure 10d**, within a matrix of 2D material. **Figures 10e** and **10f** display two different selected area electron diffractions (SAED) of sample regions A and B, respectively. These areas, can be observed as red circles in **Figure 10d**. These two distinct regions have different structural characteristics as well as diffraction patterns, which at this moment cannot be solved due to film's destruction by the electron beam and it is currently under research. In the central circle (A), the diffraction pattern discloses material of hexagonal symmetry, which could be related nanoparticles with structure resembling PbI$_2$ as it has been observed elsewhere.[47] In the last case, thus, we suggest that some unreacted PbI$_2$ or methylamine-intercalated PbI$_2$ may be in the form of undetectable units in the XRD spectra spread throughout the matrix. On the basis of the BY1 OA and PL spectra, on the other hand, there are no spectroscopic information leading to $PbI_2$ and should it exist, it has no effect on the LED device and its operation. The crucial point is that the blend is composed of an agglomerate of nanoparticles, which are assumed to reside along the n=1 nanoparticles, which are the majority species. In this manner, the observed energy transfer effects, which are notably seen as the absence of the n=1 EL, energy is transmitted towards the larger and sporadic high n nanoparticle phases.

As further test of the energy transfer model as well as the case of other related phenomena taking place in the light emission, blends with different 3D:2D molar ratio and deposition methods, were prepared. Such test films are presented in **Table 1**. The most interesting effect is the capability of having emission from the same film of two colors simultaneously, possibly due to separation of 2D (n=1) phases, each one enriched with different proportions of quasi-2D (n>1) nanoparticles which are in close proximity to the majority of 2D HOIS nanoparticles. As a result, the manufacture of a blend in the nanoscale could possibly lead to dual color led devices.

**Table 1.** Samples with different molar ratio and deposition methods.

| Sample name | Molar ratio (ZTD:2D) | Deposition method |
|---|---|---|
| **BY2** | 1:2.8 | Slow drying |
| **BY3** | 1:2.8 | Fast drying |
| **BY4** | 1:0.7 | Slow drying |
| **BY5** | 1:14 | Slow drying |

Experiments with samples described in **Table 1**, should further elaborate on the phenomena behind the absence of green light emission, due to the n=1 2D phase. Such EL phenomena are attributed to the generated excitons which transfer their energy by remote coupling to the n=2 or n=3 phases, or to the possibility that the 2D phase presents a minority not accessible by electrons and holes. OA spectra, further presented, will reveal that the 2D material is sufficient due to the strong pronounced peaks at 518 nm, however, there could be a remote possibility for some geometrical arrangement that would prohibit the n=1 phase from showing LED action. Experimental evidence will further show that even if the n=2 phase exists as a minority, it can yet funnel energy from the n=1 phase, which is achieved by remote coupling. The option in this case, of the n=2 phase sinking electrons and holes, would have required a large number of n=2 nanoparticles which are not existent as seen in the small OA peaks. Finally, experimentation has taken place in order to check if it is possible to induce EL from more than one set of specific n entities.

In **Figure 11**, OA and PL spectra of a film based on the BY2 sample, are presented. The corresponding OA spectra (**Figure 11a**) shows the existence of a n=1 phase with an OA peak at 514 nm and a n=2 phase peaked at 570 nm, while the PL spectra ($\lambda_{exc}$=450 nm) shows a slightly Stokes shifted PL peak at 527 nm. The second OA peak is broad and weak implying a wide set of n=2 like nanoparticles. As for the n=2 phase, no PL peak is observed. Performing quantitative analysis of the second low energy excitonic peak, it is assumed that the number of nanoparticles belonging to the n=2 is smaller than in the BY1 blend. **Figure 11c** shows the FF analyzed spectra, while 400 nm light is illuminating the BF. This FF PL shows the existence of a 536 nm peak, which is probably shifted w.r.t. to the peak in **Figure 11a**, due to self-absorption effects. This small shift may also be due to cavity effects coupled with interference of the excitation and emitted radiation fields. Moreover, the FF PL spectra shows a peak at 595 nm which coincides with the EL peak emission, included as an inset image in

**Figure 11**. It is important to note that the EL appears yellow from both sides of the ITO substrate, so the color emitted is not a function of self-absorption and/or re-emission, probably as it is emitted from the whole part of the sample and not just the interfaces.

In order to validate the existence of phase separation, a film based on the BY3 sample was prepared at higher temperature (110 °C), thus, much faster drying was managed. It would be expected that if a phase separation of the various n quasi-2D nanoentities existed, green color emission should be observed; such green emission is in fact readily observed from (4FpA-H)$_2$PbI$_4$ films at RT and more intensely observed at 77 K.[42] The OA and PL spectra of the aforementioned film are shown in **Figure 12**. In particular, **Figure 12a** shows a strong OA peak at 519 nm, a smaller peak at 573 nm and an almost undetectable peak at 612 nm. The PL spectra ($\lambda_{exc}$=400 nm) presented in **Figure 12b,** shows a PL peak at 526 nm while **Figure 12c** presents the PL spectra ($\lambda_{exc}$=515 nm), where a weak PL peak is exhibited at 573 nm. Moreover, the fact that only yellow light is emitted, is attributed to the high n phases neighboring to the low n phases, attract the excitonic energy and/or the carriers. It is therefore safely concluded that faster drying still allows the perception of the n=1 2D phase, but the n=2 and n=3 phase seems to be in greater abundance in comparison with the slowly dried film. This also suggests that most of the high n moieties are in close encounter to the n=1. Thus, although no PL energy is exhibited from the high order n phases, these do draw the energy in the EL mechanism, prohibiting also EL from the n=1. Of course it could be possible that the n=1 phase is not as conductive as the n=2 or n=3, due to the thinner 2D plane for n=1, which needs to be in electrical contact with the rest of the material but again such a case would be overruled by the counterargument that the pure 2D material readily shows EL.

Furthermore, images of the high methylamine content BY4 based LED while emitting, can be seen as insets of **Figure 13**. It appears that the yellow emission, ca. 592 nm is observed, although in some points the n=3 red emission is observed. The OA spectra in **Figure 13a**, shows the diminishing quantity of the n=1 phase since the OA peak at 512 nm has about the same intensity as the 536 nm peak, while both have about twice the 611 nm peak's intensity. However, in comparison to the previous OA spectra, the peaks are now much wider leading to an apparent background absorption that starts from 650 nm and downwards. The reason is that the excess of the ZTD material, transforms so much the 2D network that other phases (than the n=1,2,3) appear, which are probably crashed 2D sheets.

An interesting observation is that the BY4 sample shows red EL, at a ZTD:2D molar ratio of 1:0.7. If the light emission is attributed only to the ZTD material and the 2D does not affect the emission at all, it would be not be expected to observe strictly yellow light emission at the ratio 1:1.4.

Finally, a film based on the BY5 sample, was prepared. This specific film exhibits OA spectra (**Figure 14a**) which reveals the n=1 and n=2 phases, while the n=3 phase is barely detected. The PL spectra presented in **Figure 14b** (FF PL) shows the n=1 PL peak at 528 nm, a n=2 peak of smaller intensity at 570 nm, while the spectra of **Figure 14c** is the BF captured PL of the same film. The latter mentioned PL spectra shows the n=1 and n=3 phase only at 526 and 604 nm, respectively. It is conjectured that due to the large component of 2D n=1 phase, it is rather probable that at some points on the film the mixture may not be completely randomized w.r.t. the various n phases. However, as it is shown in **Figure 14** right inset images (various EL snapshots from the same device), yellow, green-yellow or red emission can be observed. The existence of such EL spectra means that at least 14 times more 2D material is needed to create patches of 2D n=1 phase in order to see isolated green emission from the n=1 phase. On other sample's points, the n=2 or n=3 phases do channel the excitonic energy out of the n=1 phase. Thus, emission phenomena as seen in the right inset of **Figure 14**, which are probably a complex function of dominant blends of quasi-2D and 2D phases, could readily be used for the fabrication of multi-color single layer LEDs. The left inset image shows the same LED when cooled at 77 K, where strong yellow light is observed (no spectral analysis was performed on this image to see if any green light is subsided). At last, **Figure 14c** shows the front face PL which shows minimal shift as well as a weak red peak, since the film is mainly composed of 2D phase.

As a latter presented experiment, a film of BY1 has been thoroughly rubbed in order to allow complete mixing of the various n nanoentities as well as any 3D components that exist in this composite film. The PL spectra showed clearly an enhancement of the 540 nm peak, as well as the 575 nm and the 610 nm peak. This is in accordance to the energy transfer model, where the energy is being transferred to neighboring high order n nanoentities, that have low $E_g$. This new film did not show any EL, however, LED fabrication using an active material composed of a complete randomized mix of perovskitic quasi-2D nanoentitites, is currently being pursued.

**Conclusion**

In conclusion, blend of 0D, 2D and 3D hybrid organic-inorganic semiconductor served as single layer onto LED, showed EL at low voltages. Perovskite ZTD reacted with the 2D perovskite $(4FpA-H)_2PbI_4$ (specific molar ratio) in DMF solvent and led to bright yellow light emission due to energy transfer effects, observed even with naked eye at room temperature. Device fabrication is straightforward as a single layer LED, without need of any other active layers for the emission to be accomplished.

**Supporting Information**

- Calculated XRD pattern of the 0D (MA-H)$_4$PbI$_6$ and 3D (MA-H)PbI$_3$.

- Photograph and video showing LED's yellow light emission.

**Acknowledgement**

This research did not receive any specific grant from funding agencies in the public, commercial, or not-for-profit sectors.

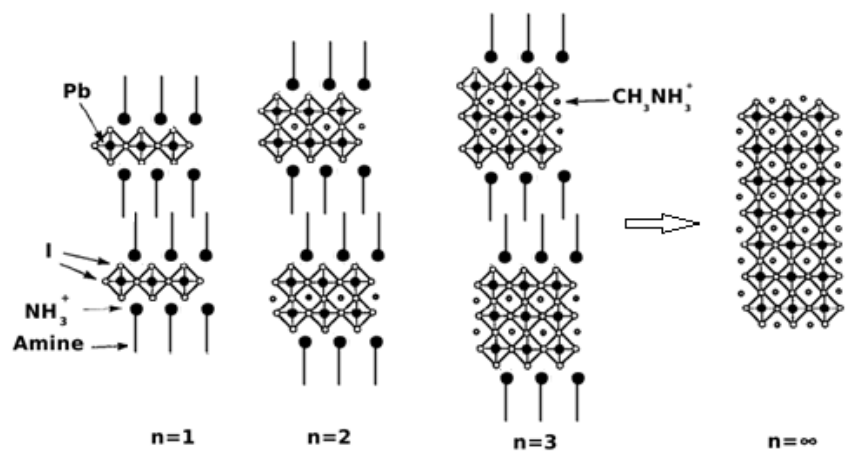

**Figure 1.** Schematic representation of the layered structure for the crystalline $(CH_3NH_3)_{n-1}$ $(4FpA-H)_2Pb_nI_{3n+1}$ LD HOIS. Dimensionalities range from 2D (left) to 3D (right), for

n=1,2,3,...,∞.

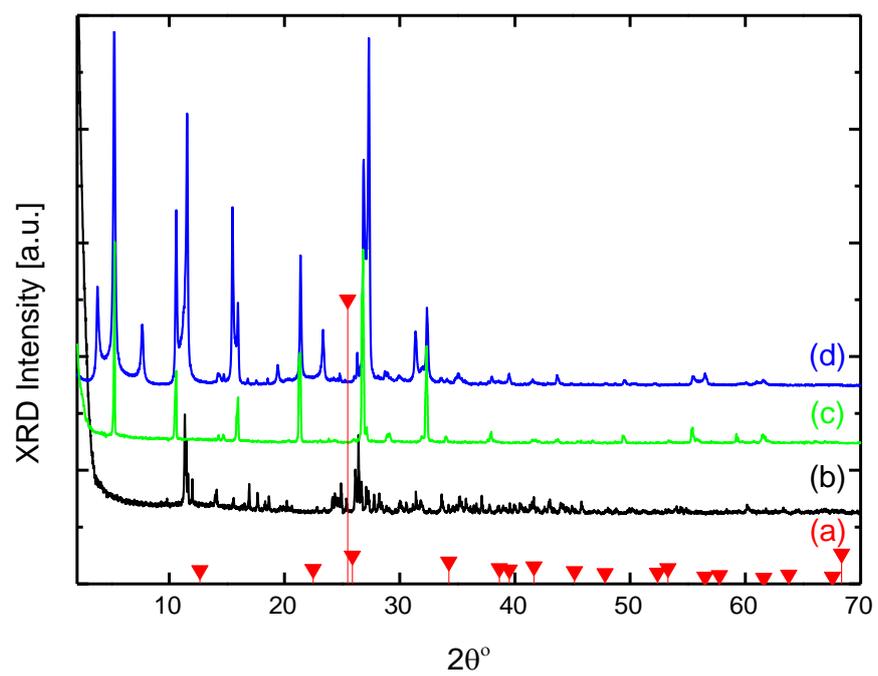

**Figure 2.** XRD patterns of (a) $PbI_2$, (b) ZTD, (c) $(4FpA-H)_2PbI_4$ and (d) BY1 blend.

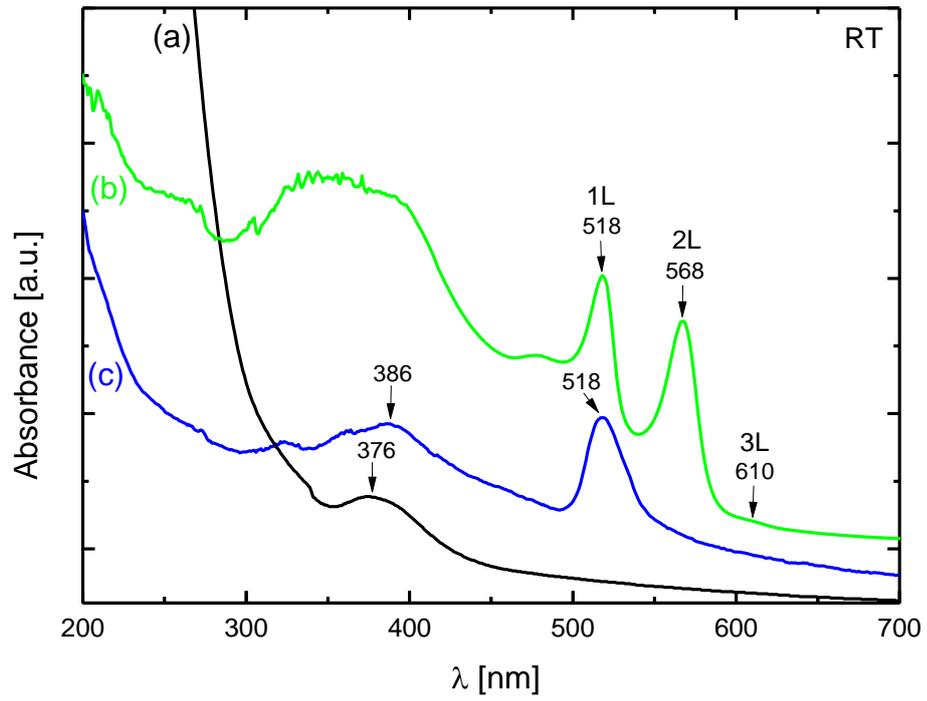

**Figure 3.** OA spectra of (a) ZTD, (b) BY1 blend and (c) (4FpA-H)$_2$PbI$_4$.

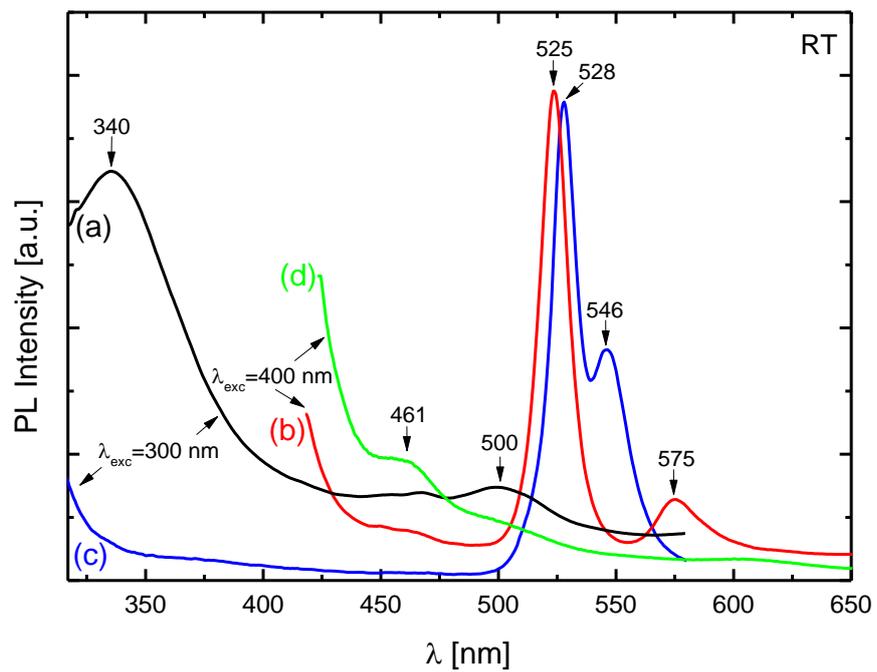

**Figure 4.** PL spectra of (a) ZTD, (b) BY1 blend, (c) (4FpA-H)$_2$PbI$_4$ and (d) ZTD. ($\lambda_{exc}$=300 nm for a,c; $\lambda_{exc}$=400 nm for b,d).

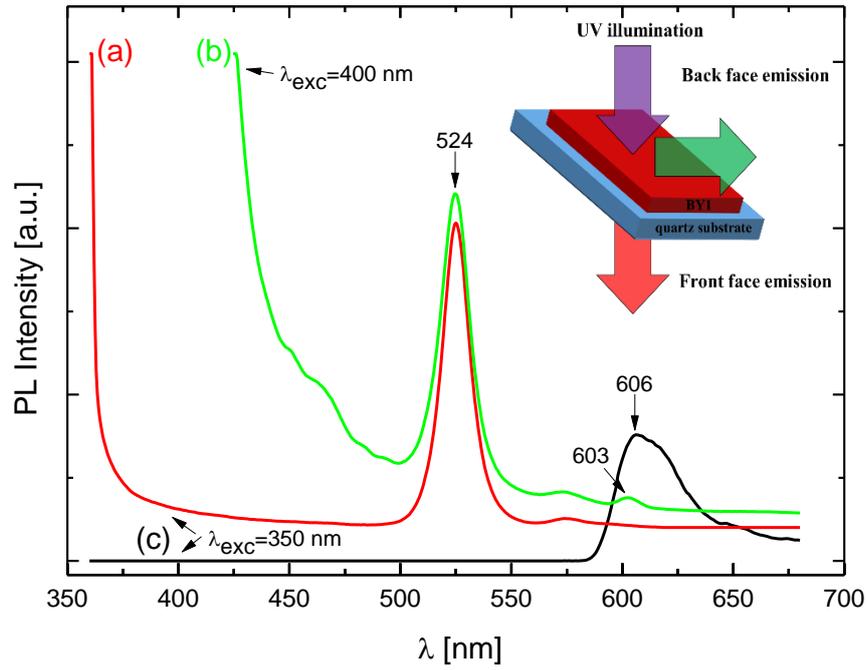

**Figure 5.** PL spectra for a BY1 film for (a) BF of a thick layer and (b) BF of the same layer swapped, (c) FF of a thin layer ($\lambda_{exc}$=350 nm for a,c; $\lambda_{exc}$=400 nm for b). Inset schematics shows PL's measurement configuration.

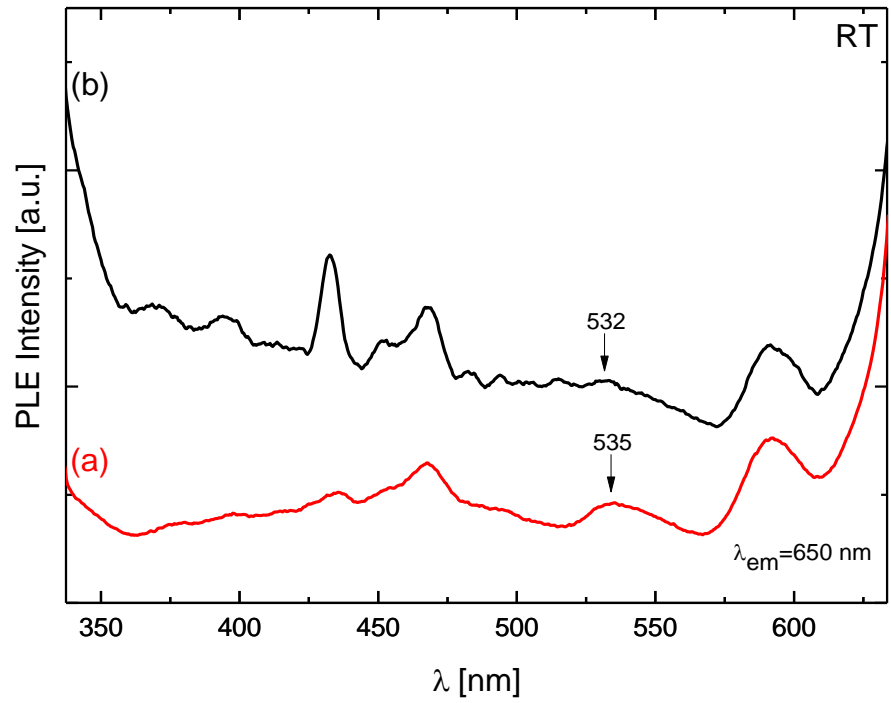

**Figure 6.** PLE spectra of BY1 ($\lambda_{em}$= 650 nm) for emission analyzed on layer's (a) BF and (b) FF, while irradiation is on the BF.

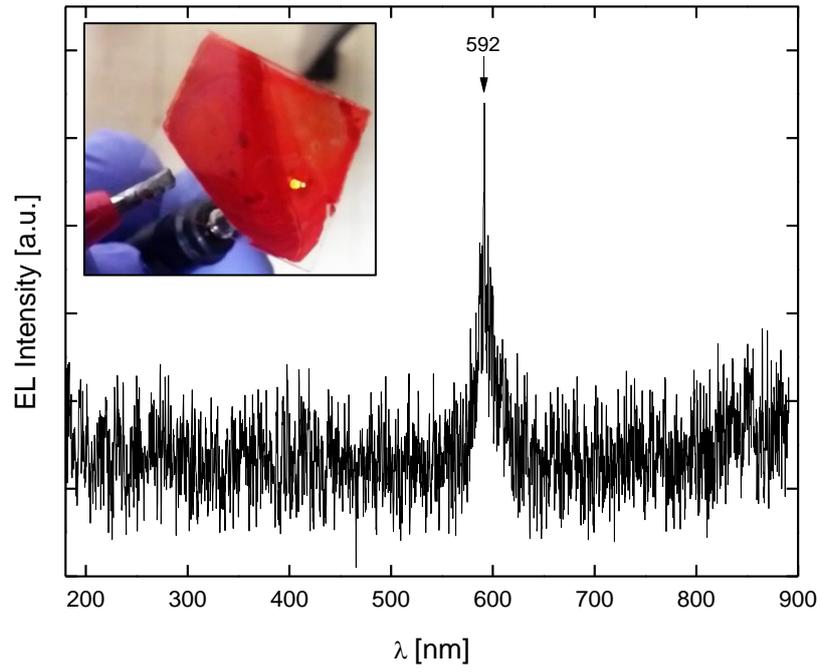

**Figure 7.** EL spectra of BY1. Inset image shows LED's yellow light emission.

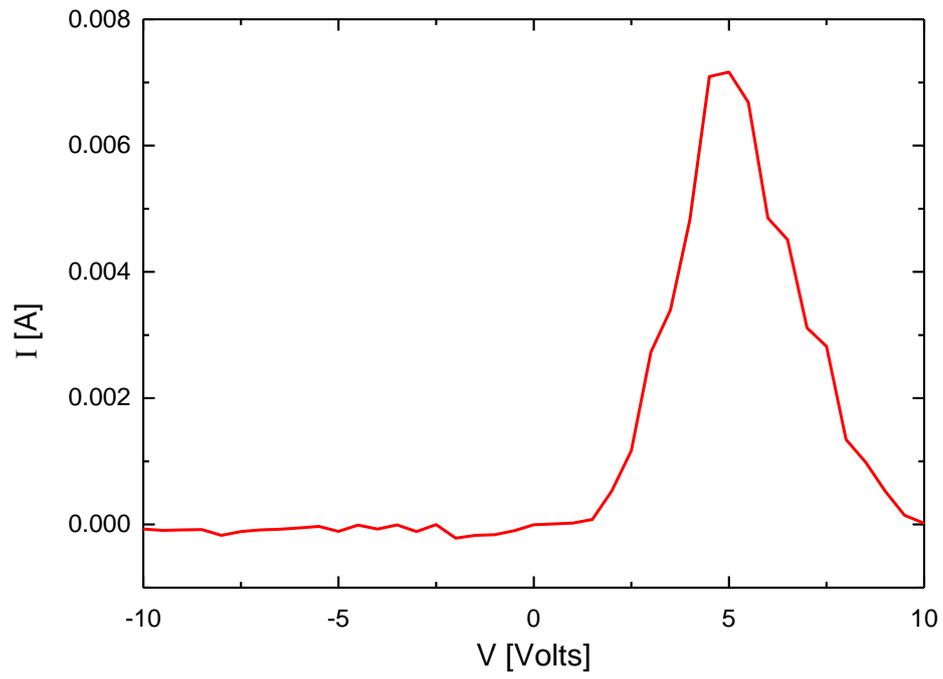

**Figure 8.** I-V characteristic curve of a LED device based on BY1.

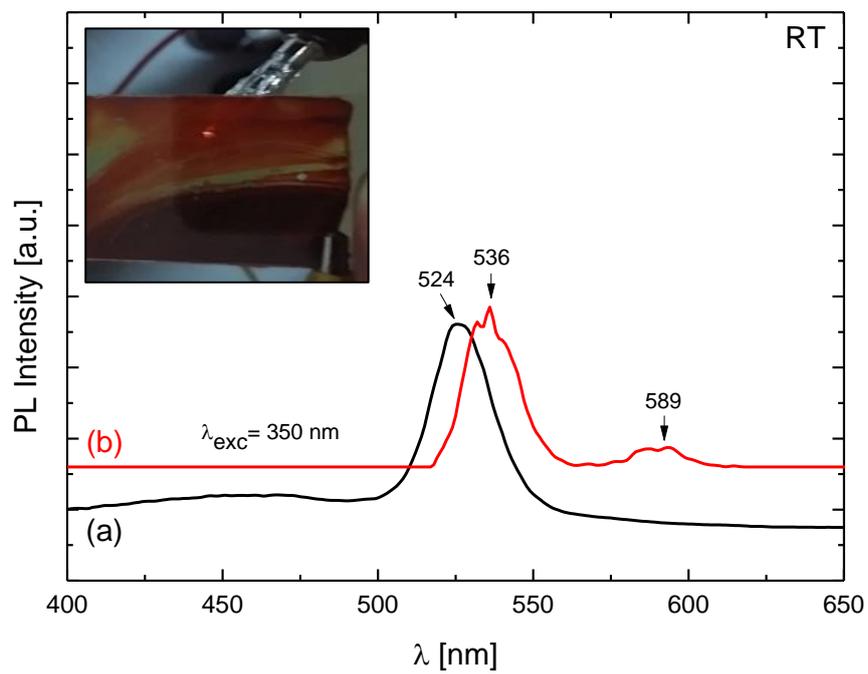

**Figure 9.** PL spectra of BY1 heated as film for 48 hours, with $\lambda_{exc}$=350 nm, for (a) BF emission and (b) FF emission. Inset shows the LED functioning after heating.

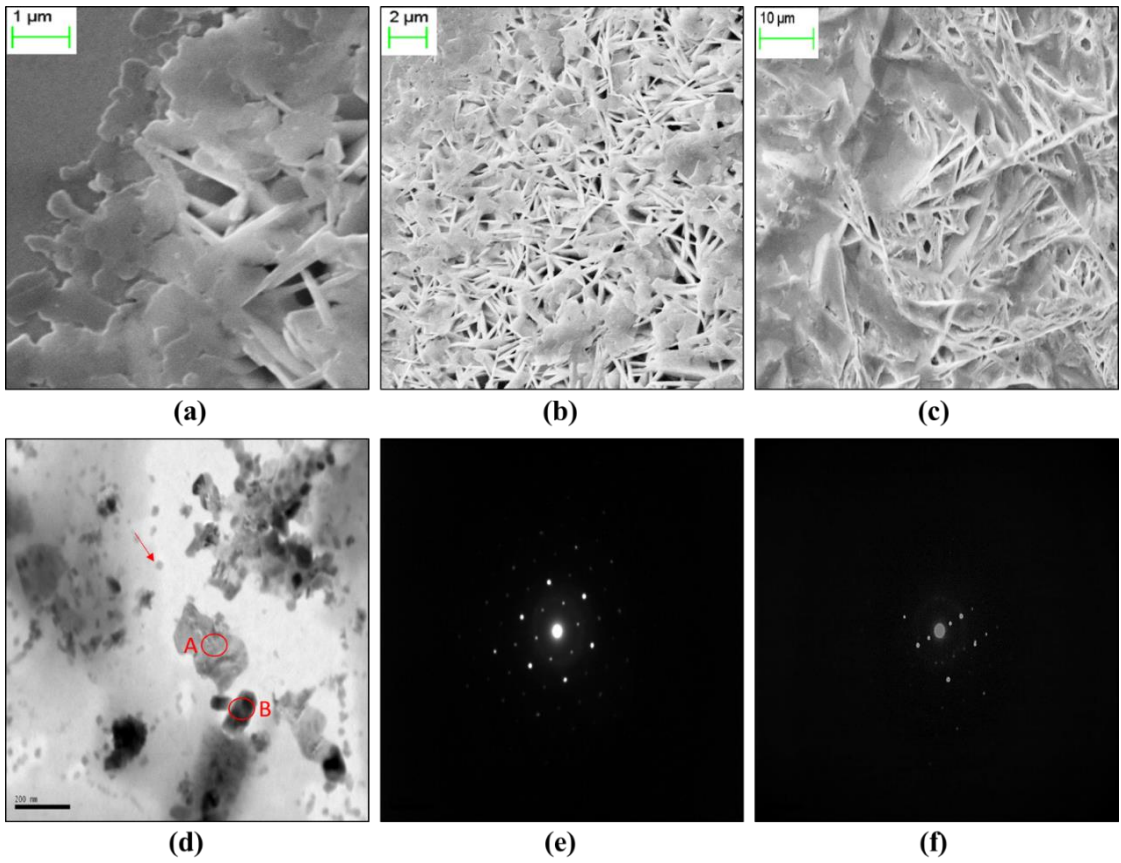

**Figure 10.** SEM (a-c) and TEM (d-f) images of BY1 blend. Images (e) and (f) represent the SAED of A and B's regions, respectively.

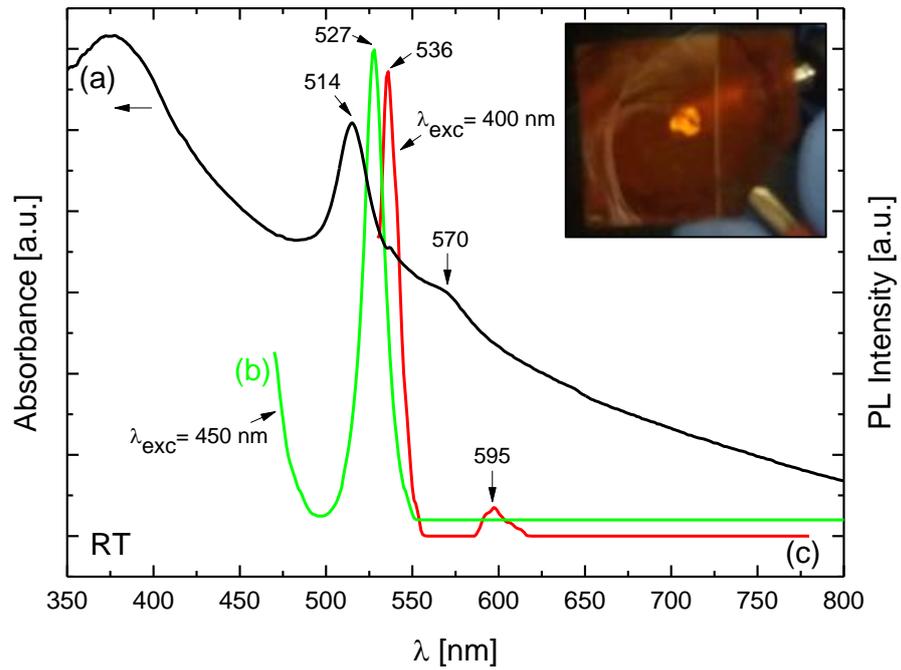

**Figure 11.** (a) OA, (b and c) PL spectra of the BY2 based film. PL is for (b) BF under $\lambda_{exc}$=450 nm and (c) FF under $\lambda_{exc}$=400 nm. Inset shows LED's emission under 8 V.

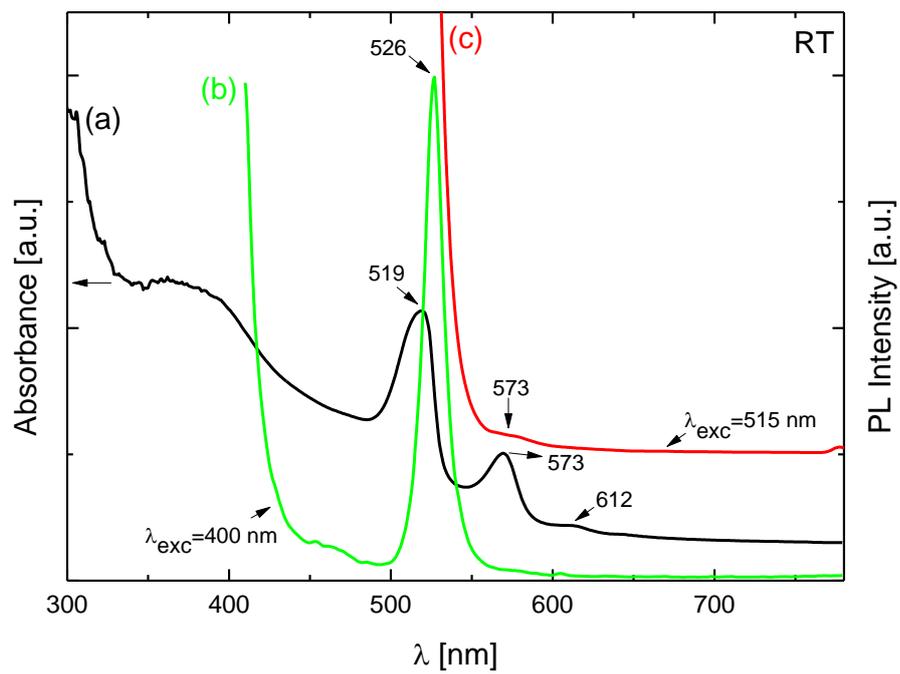

**Figure 12.** (a) OA, (b) PL ($\lambda_{exc}$=400 nm) and (c) PL ($\lambda_{exc}$=515 nm) spectra of a BY3 based film.

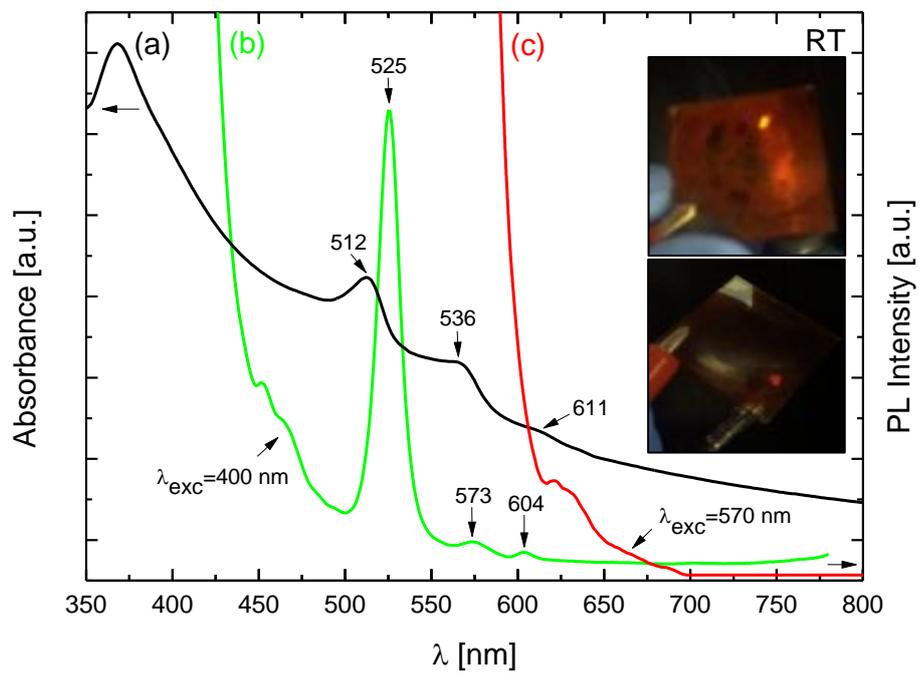

**Figure 13.** (a) OA, (b) PL ($\lambda_{exc}$=400 nm) and (c) PL ($\lambda_{exc}$=570 nm) spectra of a BY4 based film. Inset images show BY4 based LED's emission.

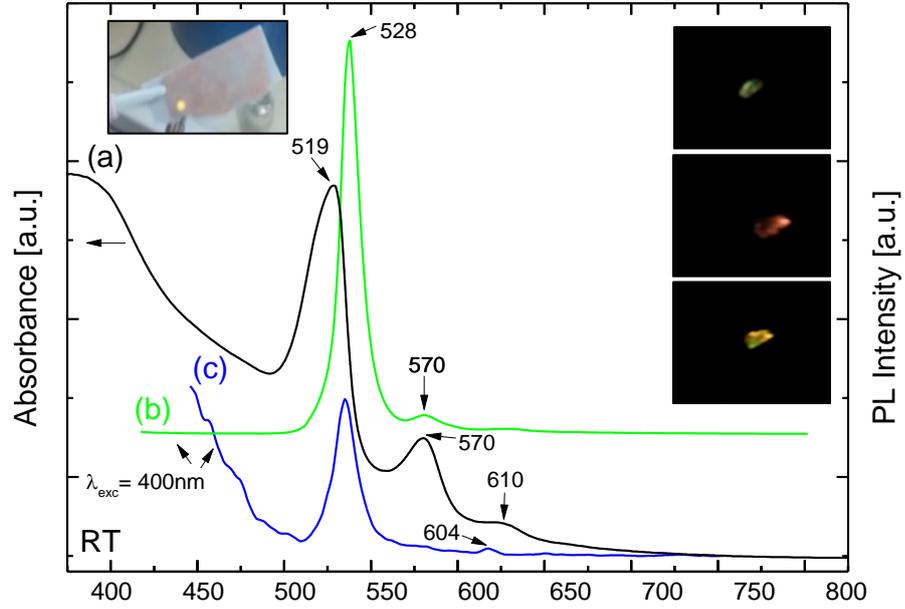

**Figure 14.** (a) OA, (b) PL for FF ($\lambda_{exc}$=400 nm) and (c) PL for BF ($\lambda_{exc}$=400 nm) spectra of a BY5 based film. Inset images: BY5 based LED's emission at 77 K (left), same LED's dual color emission at RT (right).

**TABLE OF CONTENTS**

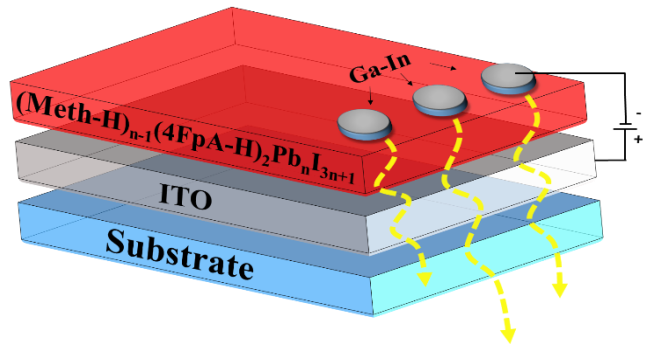

# Supporting Information

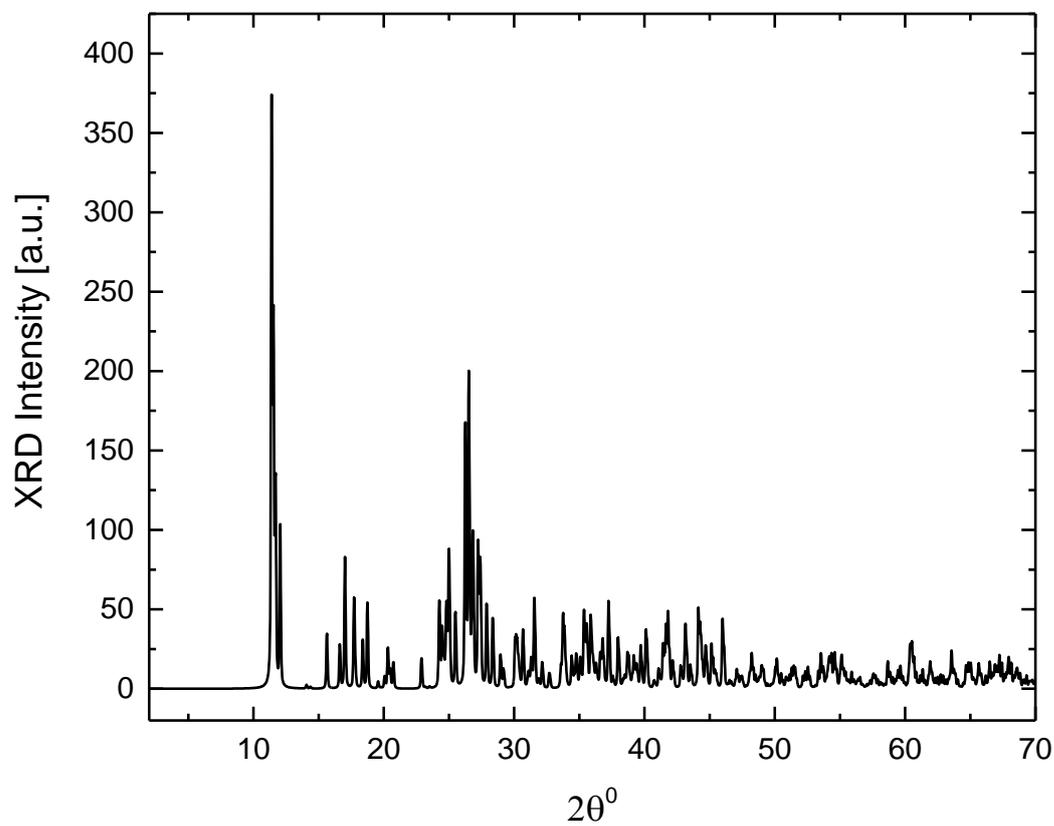

**Figure S1.** Calculated XRD pattern for the 0D (MA-H)$_4$PbI$_6$.2H$_2$O.[1]

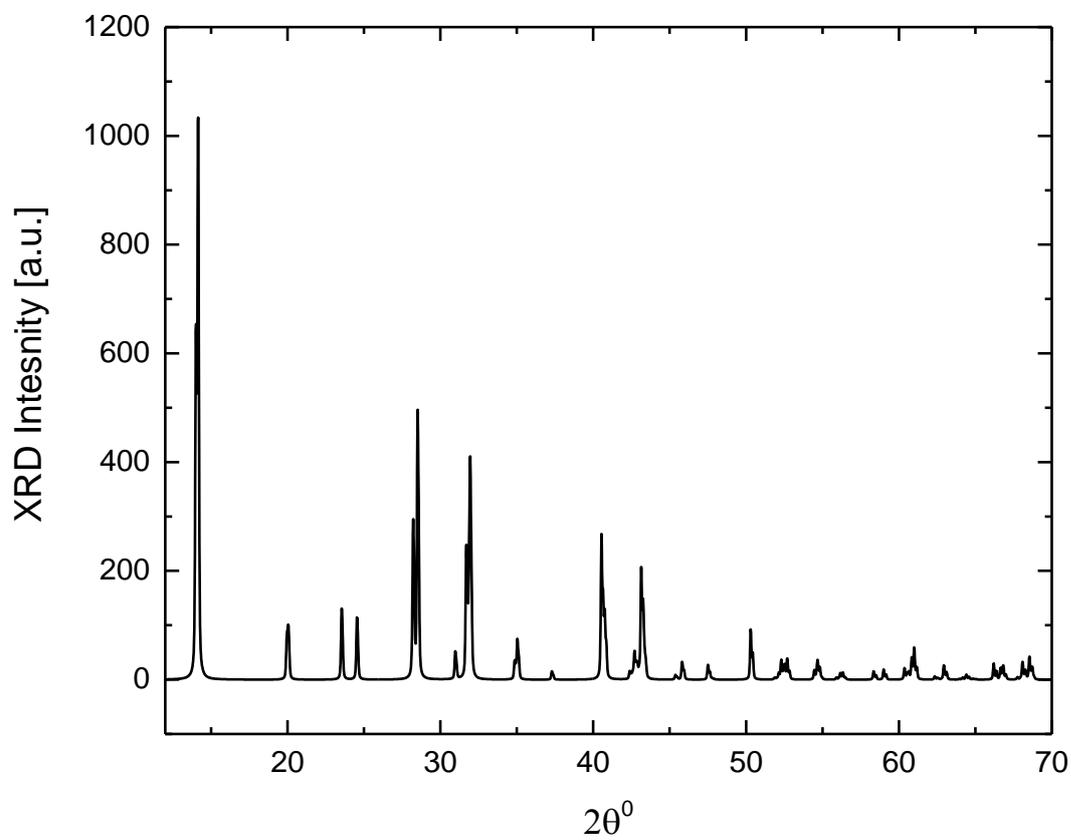

**Figure S2.** Calculated XRD pattern for the 3D (MA-H)PbI$_3$.[2]